\documentclass[aps,prl,amsmath,
twoside,floatfix,superscriptaddress,preprintnumbers]{revtex4}
\preprint{CERN-PH-TH/2005-245}
\usepackage{graphicx}
\newcommand{\bea}{\begin{eqnarray}}
\newcommand{\beq}{\begin{equation}}
\newcommand{\eea}{\end{eqnarray}}
\newcommand{\eeq}{\end{equation}}

\newcommand{\nn}{\nonumber}
\newcommand{\Frac}[2]{\frac{\displaystyle{#1}}{\displaystyle{#2}}}
\newcommand{\lsim}{\raise0.3ex\hbox{$\;<$\kern-0.75em\raise-1.1ex\hbox{$\sim\;$}}}
\newcommand{\gsim}{\raise0.3ex\hbox{$\;>$\kern-0.75em\raise-1.1ex\hbox{$\sim\;$}}}
\newcommand{\bpm}{\begin{pmatrix}}
\newcommand{\epm}{\end{pmatrix}}
\newcommand{\eq}[1]{Eq.~(\ref{#1})}
\newcommand{\unity}{{\hbox{1\kern-.8mm l}}}

\begin{document}

\title{Flavoured leptogenesis: a successful thermal leptogenesis with 
$N_1$ mass below $10^8$ GeV}
\author{O. Vives}
\affiliation{Theory Division, CERN, CH-1211, Geneva 23, Switzerland}

\begin{abstract}
We prove that taking correctly into account the lepton flavour dependence 
of the $CP$ asymmetries and washout processes, it is possible to obtain
successful thermal leptogenesis from the decays of the second right-handed
neutrino. The asymmetries in the muon and tau-flavour channels are then not
erased by the inverse decays of the lightest right-handed neutrino, $N_1$.
In this way, we reopen the possibility of ``thermal leptogenesis'' in models 
with a strong hierarchy in the right-handed Majorana masses that is typically
the case in models with up-quark--neutrino Yukawa unification. 
\end{abstract}

\maketitle
The measure of nonvanishing neutrino masses and mixings and the use of 
the seesaw mechanism to explain the smallness of these 
masses has made leptogenesis \cite{leptogenesis} one of the most promising 
mechanisms to 
generate the observed value of the baryon asymmetry of the universe.  
The seesaw mechanism requires the presence of heavy right-handed neutrinos 
with complex Yukawa couplings that can generate a lepton asymmetry through 
their out-of-equilibrium decays. This ``lepton'' asymmetry is then tranformed 
into baryon asymmetry by the standard model sphaleron processes violating 
$(B+L)$\cite{sphalerons}.  
When the right-handed neutrinos are produced through scatterings
(mainly inverse decays) in the thermal plasma this scenario is known as 
``thermal leptogenesis''.

In the thermal leptogenesis mechanism, it is usually assumed that the lepton 
asymmetry is generated by the out-of-equilibrium decays of the lightest 
right-handed neutrino, $N_1$. The asymmetry generated by the two heavier 
right-handed neutrinos, $N_2$ and $N_3$ that we take as hierarchically
heavier than $N_1$, is usually assumed to be either negligible or efficiently  
washed-out by the $N_1$ inverse decays \cite{dibaribuchplum}. In these 
conditions only $N_1$ generates a baryon asymmetry.
The final value of the baryon asymmetry normalized to the entropy density, 
$\eta_{\rm B}$, is then related to the $CP$ asymmetry, $\varepsilon_1$, in 
the decays of the lightest right-handed  neutrino as
\bea
\label{etab}
\eta_{\rm B} = \Frac{8}{23} ~\Frac{n_{N_1}}{s} ~ \chi_1~ 
\varepsilon_1 \simeq 3 \times 10^{-4} ~ \chi_1~ \varepsilon_1, 
\eea
where the factor $8/23$ is the fraction of the $B-L$ asymmetry
converted into baryon asymmetry by sphalerons. $n_{N_1}/s$  is the number
density of right-handed neutrinos normalized to the entropy density that in
equilibrium is approximately equal to $0.2/g_*$ with $g_*\simeq 230$ 
the number of propagating states in the supersymmetric plasma. Finally 
$\chi_1$ is the efficiency  factor describing the 
fraction of the $CP$ asymmetry that survives the washout by inverse decays and
scattering processes. Now, for hierarchical right-handed neutrinos, it is 
possible to relate $\varepsilon_1$ to the mass of the lightest right-handed
neutrino, the observed neutrino masses, and an effective leptogenesis phase as
\bea
\label{epsilon}
\varepsilon_1 = \Frac{3}{8 \pi} ~ \Frac{M_1}{v_2^2} \left(m_3 - m_1 \right) ~
\sin \delta_{\rm L} \leq \Frac{3}{8 \pi} ~ \Frac{M_1~m_{\rm atm}}{v_2^2} 
\simeq 2 \times 10^{-6} \left(\Frac{M_1}{10^{10}~ {\rm GeV}}\right)\left(\Frac{
m_{\rm atm}}{0.05~ {\rm eV}}\right),
\eea    
where the maximum value for this asymmetry corresponds to fully hierarchical
neutrinos with $m_3 = m_{\rm atm} \simeq 0.05$ eV, $m_1\simeq 0$ and a
maximal phase $\delta_{\rm L}$.

The baryon asymmetry is precisely measured
from the cosmic microwave background as $\eta_{\rm B}^{\rm CMB} = 
0.9 \times 10^{-10}$ \cite{CMB}. Thus, using Eqs.~(\ref{etab}) and 
(\ref{epsilon}) with $\chi_1 =1$ we obtain a lower bound on the mass of $N_1$
\cite{sachaalej}: 
\bea
\label{bound}
M_1 \geq M_1^{\rm min} = 1.5 \times 10^9~{\rm GeV} \left(\Frac
{\eta_{\rm B}^{\rm CMB}}{9 \times 10^{-11}}\right)
\left(\Frac{0.05~ {\rm eV}}{m_{\rm atm}}\right).
\eea
However this lower bound represents a serious problem for many
flavour models, specially those incorporating some grand unification symmetry.
In these models we usually expect that neutrino Yukawas are closely 
related to the up-quark Yukawa matrices. As show in the
appendix of \cite{sacha} (see also Ref. \cite{branco}) and explained below, 
in these models it is possible 
to obtain the mass of the lightest right-handed neutrino as
\bea 
\left|M_1\right| \simeq \Frac{y_1^2 v_2^2}{\left|W_{1 k}^2~ m_{\nu_k}\right|} 
\simeq \Frac{m_u^2}{\left|W_{11}^2~ m_1 + W_{12}^2~ m_{\rm sol} + 
W_{13}^2 ~m_{\rm atm}\right|},
\eea
with $W_{ij}$ the rotation from the basis of diagonal $\nu_{\rm L}$ masses to
the basis where the (left-handed) neutrino Yukawa matrix $Y_\nu Y_\nu^\dagger$
is diagonal, $m_{\nu_k}$ the left-handed neutrino masses that we take as 
hierarchical and we have approximated $y_1 v_2 \simeq m_u$. From this 
expression 
it is straightforward to see that we can only obtain $M_1 \geq 10^9~{\rm GeV}$
if
\bea
 \Frac{m_u^2}{10^9~{\rm GeV}} = 4 \times 10^{-6}~{\rm eV}
\left(\frac{m_u^2}{\left(2~{\rm MeV}\right)^2}\right) \gsim
\left|W_{11}^2~ m_1 + W_{12}^2~ m_{\rm sol} + W_{13}^2~ m_{\rm atm}\right|.
\eea
This implies that, barring accidental cancellations (or some specific textures
as in \cite{strumiamarti}), having a large enough $N_1$ {\bf is only possible}
if we have simultaneously $m_1 \lsim  4 \times 10^{-6}~{\rm eV}$, 
$W_{12}^2\lsim 5 \times 10^{-4}$ (for $m_{\rm sol}\simeq 0.008$ eV) and 
$W_{13}^2 \lsim 8  \times 10^{-5}$ (for $m_{\rm atm}\simeq 0.05$ eV).
These conditions are usually not satisfied in most flavour models where
the typical values for the lightest right-handed neutrino mass are close to 
$10^6$--$10^7$ GeV with $W_{12}$ of the order of mixing angle in solar 
neutrinos \cite{su3,models,double}. Therefore, apparently, these models 
can not produce a sufficient baryon asymmetry through the thermal leptogenesis 
mechanism.   

It is evident that this represents a very serious problem for all these
flavour models, that are forced to abandon thermal leptogenesis and use
other mechanisms to produce the observed baryon asymmetry \cite{other}. 
In this paper we 
present a solution to this problem. We show that when the {\rm flavour 
dependence} of the lepton asymmetries and the erasure processes are 
correctly taken into account, it is still possible to generate a large enough
baryon asymmetry in the decays of the second right-handed neutrino $N_2$
that then survives the washout due to the inverse decays of $N_1$.
The basic idea behind this statement is that the asymmetries generated by $N_2$
in the different flavour channels are washed out by different $N_1$ Yukawa
interactions.  In fact, the decays of the second
right-handed neutrino $N_2$ create different asymmetries in the $\tau$, 
$\mu$ and $e$ channels. As we will show these asymmetries do not mix through
the thermal interactions with the plasma before the $N_1$ inverse decays 
become effective. Therefore, it is evident that the different flavour channels
will only be erased by the corresponding $N_1 \to H L_i$ interaction and not
by the sum over all lepton flavours as usually done
\cite{dibaribuchplum,sacha,diBari,plumacher,pedestrians,thermal}. 
Taking into account these effects, 
we prove here that in models with  $Y_\nu\simeq Y_u$, the lepton asymmetry in
the $\tau$ and $\mu$ channels is only mildly erased by $N_1$ decays.  
Some of these ideas were already present in the literature. 
For instance, the possibility of generating the asymmetry in 
$N_2$ decays was previously discussed by P. di Bari in the one flavour 
approximation \cite{diBari}; an early discussion of flavour effects in 
leptogenesis, mainly in $N_1$ decays, can be found in \cite{BCST}, and 
flavour effects in the context of low scale leptogenesis models were
recently studied in \cite{underwood}. In this paper we show that combining 
these elements it is possible to obtain successful leptogenesis in realistic
models of fermion masses with up-quark--neutrino Yukawa  
unification. Therefore, we reopen the possibility of 
``thermal leptogenesis'' in these models.

In the following we give a general proof of this statement.
Although this mechanism is equally valid in supersymmetric and
non-supersymmetric models we only discuss the non-supersymmetric
example. The supersymmetric case is completely analogous \cite{susy}.

The general expression for leptonic Yukawa couplings and Majorana masses in a
seesaw Standard Model is, ${\cal L}_{\rm Yuk}~=~ L^T Y^e e_{\rm R}^c~ H_1 
~+~ L^T Y^\nu 
\nu_{\rm R}^c~ H_2 ~-~ \frac{1}{2}~{\nu_{\rm R}^c}^T {\cal M}\nu_{\rm R}^c$,
with $L_i, e_{R\,i},\nu_{R\,i}$, $(i=e,\mu,\tau)$ the three generations of 
leptons and $Y^e$, $Y^\nu$ and ${\cal M}$, $3 \times 3$ matrices. In the basis
of diagonal $Y^e$ and ${\cal M}$, the neutrino Yukawa matrix is $Y^\nu = 
V_{\rm L}^\dagger\cdot D_{Y_\nu}\cdot V_{\rm R}$, where $D_{Y_\nu}$ is a 
diagonal mass of neutrino Yukawa couplings, that in the class of models
discussed here is $D_{Y_\nu} = {\rm diag}(y_1,y_2,y_3) \simeq 
{\rm diag}(m_u/v_2,m_c/v_2,m_t/v_2)$. The effective Majorana mass for the 
left-handed neutrinos is obtained through the seesaw mechanism as
\bea 
\label{seesawmass}
m_\nu~ =~ v_2^2~ {Y^\nu}\cdot {\cal M}^{-1} \cdot {Y^\nu}^T~ =~ U \cdot 
D_{m_\nu} \cdot U^T,
\eea
where the mixing matrix $U$ is the  Pontecorvo-Maki-Nakagawa-Sakata (PMNS) 
matrix that is sufficiently known from neutrino
experiments, and $D_{m_\nu}= {\rm diag}(m_1,m_2,m_3)$ the three
neutrino masses. At the moment, we know only two mass differences, 
$\Delta m^2_{\rm atm} \simeq 2.7 \times
10^{-3}~{\rm eV}^2$ and $\Delta m^2_{\rm atm} \simeq 7.0 \times
10^{-5}~{\rm eV}^2$. Although in the following we assume hierarchical 
spectrum with normal hierarchy, other cases can be easily treated in the same 
way.

Now we define the hermitian matrix $\Lambda\equiv v_2^4 y_1^4~
{\cal M}^{-1\,\dagger} {\cal M}^{-1}$ \cite{sacha} and, in the basis of 
diagonal $Y^{\nu\,\dagger} Y^\nu$, from 
\eq{seesawmass} we have
\bea
\frac{\Lambda}{v_2^4 y_1^4} = D_{Y_\nu}^{-1}~ V_{\rm L}^*~ m_\nu^\dagger~ 
V_{\rm L}^\dagger~
D_{Y_\nu}^{-2}~ V_{\rm L}~ m_\nu ~V_{\rm L}^T~D_{Y_\nu}^{-1}~~ \equiv~~ 
D_{Y_\nu}^{-1}~ \Delta^\dagger~D_{Y_\nu}^{-2}~\Delta~D_{Y_\nu}^{-1},
\eea
with $ \Delta = V_{\rm L}~ m_\nu~V_{\rm L}^T \equiv W~D_{m_\nu}~ W^T$. From 
here the matrix $\Lambda$ can be written as
\bea
\Lambda_{ij} = \left( \vec \lambda_i^\dagger\right)\cdot 
\left( \vec \lambda_j\right) \equiv 
\sum_k\left(\vec \lambda_i^*\right)_k
\left( \vec \lambda_j\right)_k, \qquad
\vec \lambda_i = \frac{y_1}{y_i} \bpm \Delta_{1i}\\\frac{y_1}{y_2} 
\Delta_{2i}\\ \frac{y_1}{y_3} \Delta_{3i} \epm.
\eea
Given the strong hierarchy in the neutrino-Yukawa eigenvalues (similar to the
hierarchy in the up-quark masses) we have in practice always that
$|\vec \lambda_1|^2 \gg|\vec \lambda_2|^2, |\vec \lambda_3|^2$. For instance,
to change this hierarchy in the simplified case where we take $W_{13}\simeq 0$ 
and $y_1/y_3\simeq0$ would require that $|\Delta_{11}| < |\Delta_{22}| 
y_1^2/y^2$.
In these conditions it is very easy to understand that the largest eigenvalue
of the matrix $\Lambda$, corresponding to $v_2^4 y_1^4/|M_1|^2$, will be  
given in very good approximation by the $(1,1)$ element of this matrix. 
Therefore the lightest right-handed Majorana neutrino mass is given by: 
\bea
\label{M1}
|M_1| \simeq \Frac{y_1^2 v_2^2}{|\lambda_1|} \simeq  \Frac{y_1^2 v_2^2}{|
\Delta_{11}|} = \Frac{m_u^2}{|W_{1k}^2 m_{\nu_k}|}.
\eea
From this expression we can estimate the mass of the lightest right-handed 
neutrino. If the observed neutrino mixings are not present in the 
neutrino-Yukawa matrices and come mainly from the seesaw mechanism itself 
we have that
$V_{\rm L} \simeq \unity$ and $W\simeq U_{\rm PMNS}$. Then we have that
$|M_1| \simeq 2~m_u^2/m_{\rm sol} \simeq 1 \times 10^6$ GeV. Only  in the case
$W\simeq\unity$ we can have $|M_1| \simeq m_u^2/m_1= 10^8 
\left(4 \times 10^{-5} {\rm eV}/m_1\right)$ GeV. Therefore, typically $N_1$ 
is too light to generate a sizeable baryon asymmetry. 
In the same way the associated eigenvector to this eigenvalue is given by 
$\hat \lambda_1 \simeq \vec \lambda_1/\Delta_{11}$.
Using this eigenvector we can also obtain the Yukawa couplings of 
$N_1$ in the basis of diagonal $Y^e$ and ${\cal M}$ as
\bea
\label{N1yuk}
\left(Y_\nu\right)_{i 1} = \left(V_{{\rm L}}^\dagger \cdot D_{Y_\nu} 
\left(\hat\lambda_1 \right)\right)_i = \frac{y_1}{\Delta_{11}}~ 
V_{{\rm L}}^\dagger \cdot \bpm \Delta_{11} \\ \Delta_{21} \\ 
\Delta_{31} \epm = \frac{y_1}{\Delta_{11}}~  \cdot \bpm  U_{1k}~ 
m_{\nu_k}~ W_{1k} \\ U_{2k}~ m_{\nu_k}~ W_{1k} \\ U_{3k}~ m_{\nu_k}~ 
W_{1k} \epm,
\eea
where we used that $ \sum_i \left(V_{{\rm L}}\right)_{i j}^* \Delta_{i1} =
\sum_k U_{jk} m_{\nu_k}  W_{1k}$.

At this point we can already discuss the behaviour of a lepton asymmetry
created by the out-of-equilibrium decays of the next-to-lightest right-handed
neutrino $N_2$. The asymmetry in the different flavour channels 
$(a=e,\mu,\tau)$, assuming $M_1 \ll M_2 \ll M_3$, is given by \cite{diBari}
\bea 
\label{asym2}
\epsilon_2^a = \Frac{1}{8 \pi \left(Y^\dagger Y\right)_{22}} \left(\frac{3}{2}
~\frac{M_2}{M_3}~ {\rm Im} \left[Y_{a 2}^* Y_{a 3} Y_{k 2}^* Y_{k 3} \right]
+\frac{M_1}{M_2} ~{\rm Im} \left[Y_{a 2}^* Y_{a 1} Y_{k 2}^* Y_{k 1} \right]
\right). 
\eea
Owing to the hierarchy in the neutrino Yukawa matrices and the 
right-handed Majorana neutrinos, the first term in \eq{asym2} usually 
dominates and therefore we take it to estimate the size of the asymmetry.  
Now we have that, $1/v_2^2~{\rm Im} \left[ Y_{a 2}^* \left(m_\nu\right)_{a k} 
Y_{k 2}^*\right] = 1/M_3~{\rm Im} 
\left[Y_{a 2}^* Y_{a 3} Y_{k 2}^* Y_{k 3} \right] + 1/M_1~{\rm Im} 
\left[Y_{a 2}^* Y_{a 1} Y_{k 2}^* Y_{k 1} \right]$.
Therefore, the numerator in $\epsilon_2^a$ includes the $M_3$ contribution 
to the $\left(m_\nu\right)_{a k}$ element. To maximize the asymmetry we 
assume this element to be of order $m_{\nu_3} = m_{\rm atm}$ and define
the corresponding $CP$-violating phase as $\delta_\nu$, then we have
\bea
\label{estimate}
\epsilon_2^a \simeq \Frac{3}{8 \pi} ~\frac{M_2 ~ m_3}{v_2^2}~ 
\Frac{|Y_{a 2}|}{|Y_{3 2}|}~ \delta_\nu,
\eea
where we use that the hierarchy in the Yukawa elements is such that
$Y_{3 j} > Y_{2 j}, Y_{1 j}$. 
It is clear that asymmetry in the $\tau$ channel can be as 
large as $\epsilon_2^\tau = 3 \times 10^{-6} (M_2/10^{10}~ {\rm GeV})$. In 
the muon and electron channels the asymmetry will be further suppressed by
$|Y_{2 2}|/|Y_{3 2}|$ and $|Y_{1 2}|/|Y_{3 2}|$ respectively, and
thus we can expect a smaller asymmetry specially in the electron channel
\footnote{In many flavour models $|Y_{2 2}|\simeq|Y_{3 2}|$ to explain the
  relation between $m_s/m_b$ and the CKM element $V_{cb}$, although this is
  not necessary in the up sector \cite{RRRV}.}.

These asymmetries are partially converted, through sphaleron processes, to a
$\Delta_a = \frac{1}{3}B-L_a$ asymmetry and, in part, washed away by $N_2$ 
inverse decays. The final lepton asymmetry after the decay of $N_2$ will be 
given by
\bea
\eta_{\Delta_a}~ \simeq~ \frac{n_{N_{2}}}{s}~ \epsilon_2^a ~ 
\chi_2^a 
\eea
with $n_{N_{2}}/s$ the number density of $N_2$ normalized to the
entropy density\footnote{Strictly speaking
we should include a $3 \times 3$ matrix $A$ connecting the lepton asymmetries
in the different lepton flavours $L_a$ to the $\Delta_a$ asymmetries 
\cite{BCST}. However, at $T> 10^9$ GeV these constant matrices are very close 
to the identity and their inclusion does not change the results below}. 
The efficiency factor,
$\chi_2^a$, is the fraction of the produced asymmetry that survives 
after the end of $N_2$ decays and inverse decays and it is approximately 
given by
\bea
\chi_2^a \simeq \Frac{\Gamma(N_2\to l_a H)}
{\left. H\right|_{T\simeq M_2}}
\simeq \Frac{\tilde m_2^a}{m_*},
\eea
with $m_2^a \equiv v_2^2~ Y_{a 2}Y_{a 2}^*/M_2$ \cite{plumacher} and 
$m_*$ is the equilibrium 
mass equal to $\simeq 1 \times 10^{-3}$ eV in the SM \cite{plumacher}. 
The final 
values of this efficiency factor are model dependent and therefore we keep 
this factor as a parameter in the following.

At temperatures slightly  below $M_2$ we have three different asymmetries 
in the $\Delta_\tau$, $\Delta_\mu$ and $\Delta_e$ channels. These asymmetries 
remain basically unchanged from $T\sim M_2$ to  $T\sim M_1$ and, in
particular, the different flavours do not mix. In fact, at this temperature,
both $N_2$ and $N_3$ right-handed neutrinos have already decayed. Therefore
the only possible Yukawa interactions in the plasma are those with the 
right-handed charged leptons and with $N_1$. Taking into account that the 
$Y^e_{3,3}$ Yukawa interactions are in thermal equilibrium at temperatures 
below $\simeq 10^{14}$ GeV, it is more convenient to use 
the basis of diagonal charged lepton Yukawas to include these interactions 
in the thermal masses. Thus, only the Yukawa interactions with $N_1$ can 
change flavour, although $N_1$ are very rare in the plasma up to temperatures 
of the order of $M_1$ when they reach the equilibrium density 
\cite{pedestrians,thermal}.
Therefore, we have to discus the washout of the different asymmetries by the
$N_1$ decays and inverse decays. First, it is easy to see that this washout 
will be exponential and not linear if the $N_1$ are thermally produced at
temperatures close to $M_1$. From Ref. \cite{pedestrians} and taking 
$\varepsilon_1\simeq 0$, the Boltzman equation for this case would be
\footnote{Notice that, due to the exponential washout, order one factors in
the $A$ matrices can be important \cite{BCST}.}:
\bea
\label{n1wash}
\Frac{\partial \eta_{\Delta_a}}{\partial z} = - \frac{1}{4} z^2 e^{-z}
\sqrt{1 + \frac{\pi}{2} z}~ \Frac{\tilde 
m_1^a}{m_*}~ A_{ab}~ \eta_{\Delta_b}, \qquad {\rm with} 
\quad
A (T \leq 10^9 {\rm GeV}) \simeq \bpm -0.86 & 0.1 & 0.1 \\ 0.06 & - 0.65 & 0.017
\\0.06&0.017&-0.65 \epm,
\eea
with $\tilde m_1^a\equiv v_2^2~Y_{a 1}Y_{a 1}^*/M_1$ takes into account that 
a given asymmetry $\Delta_a$ can only be erased by the corresponding lepton 
flavour, $z=M_1/T$ and $m_* \simeq 1 \times 10^{-3}$ eV\footnote{In this
  equation we consider only the effects of decays and inverse decays. The
  effects of $\Delta L=1$ and $\Delta L=2$ scatterings do not change the
  results presented here for $\tilde m_1^a \geq 10^{-3}$ eV 
  \cite{pedestrians}.}. Now we can use 
Eqs.~(\ref{M1}) and (\ref{N1yuk}) and we see that
\bea
\label{washout}
\tilde m_1^e& =& |\sum_k U_{1k} m_{\nu_k} W_{1k}|^2/
|\Delta_{11}| \simeq | m_{2} W_{12}/\sqrt{2} +
m_{1} W_{11}/\sqrt{2}|^2/|\Delta_{11}| \nn \\
\tilde m_1^\mu &=& |\sum_k U_{2k} m_{\nu_k} W_{1k}|^2/
|\Delta_{11}|  \simeq | - m_{2} W_{12}/ 2 +
m_{1} W_{11}/2 + m_{3} W_{13}/ \sqrt{2}|^2/|\Delta_{11}|  \nn \\
\tilde m_1^\tau &=& |\sum_k U_{3k} m_{\nu_k} W_{1k}|^2/
|\Delta_{11}| \simeq | m_{2} W_{12}/2 -
m_{1} W_{11}/2 + m_{3} W_{13}/ \sqrt{2}|^2/|\Delta_{11}|, 
\eea
where $m_3 = m_{\rm atm}$ and $m_3 = m_{\rm sol}$ and we have used bimaximal 
mixings in the PMNS matrix to make an estimate. The final values of 
$\tilde m_1^a$ depend on the matrix $W$. In case the case the mixings in the
neutrino Yukawa matrix are small, we have that $W \simeq U$ and so
\bea
\label{WeqU}
\tilde m_1^e \simeq  \left|\frac{m_{2}}{2} +\frac{m_{1}}{2}\right| \simeq 
\Frac{m_{\rm sol}}{2},~\quad 
\tilde m_1^\mu \simeq  \left|-\frac{m_{2}}{2\sqrt{2}} 
+\frac{m_{1}}{2\sqrt{2}}\right|^2/\left|\frac{m_{2}}{2}\right| 
\simeq \Frac{m_{\rm sol}}{4},~ \quad
\tilde m_1^\tau \simeq  \left|\frac{m_{2}}{2\sqrt{2}} 
+\frac{m_{1}}{2\sqrt{2}}\right|^2/\left|\frac{m_{2}}{2}\right| 
\simeq \Frac{m_{\rm sol}}{4}.~~ 
\eea
Using these values in the case $\eta^{\rm ini}_{\Delta_\tau} \simeq 
\eta^{\rm ini}_{\Delta_\mu} \gg \eta^{\rm ini}_{\Delta_e}$, \eq{n1wash} 
can be easily integrated (taking $\eta_{\Delta_\tau}(z)\simeq 
\eta_{\Delta_\mu}(z)$, $\eta_{\Delta_e}(z)\simeq 0$ and 
$A_{\tau\tau} + A_{\tau\mu}=0.63$) and we obtain
\bea
\eta^{\rm fin}_{\Delta_{\mu,\tau}} = \eta^{\rm ini}_{\Delta_{\mu,\tau}} ~ 
e^{-0.75 ~\tilde m_1^{\mu,\tau}/m_*}.
\eea
If now we take the central value from the solar mass difference as 
$m_{\rm sol} = 0.008$ eV, we have that the muon and tau 
asymmetries are only erased by a factor  ${\rm exp}(-1.5) \simeq 0.22$ and
hence they can survive the washout from $N_1$ decays. This should be compared 
with the washout that we would obtain in the single flavour 
approximation, that would be $\simeq e^{-0.75 ~\sum_a \tilde m_1^a/m_*} =
e^{-0.75~m_{\rm sol}/m_*} = 2.4 \times 10^{-3}$. 
Now the final asymmetry in the $\tau$ channel would be
\bea
\eta^{\rm fin}_{\Delta_\tau} &\simeq& 0.001 \chi^\tau_2 ~\Frac{3}{8 \pi} 
~\frac{M_2 ~ m_3}{v_2^2}~ \delta_\nu~ e^{-0.75~ m_{\rm sol}/(4 m_*)}
\lsim  6.5 \times 10^{-10} (M_2/10^{10}~ {\rm GeV})~\chi^\tau_2,\\
\eta_B &=& \Frac{28}{79} \left(\eta^{\rm fin}_{\Delta_\tau} + 
\eta^{\rm fin}_{\Delta_\mu} \right) \simeq \Frac{56}{79}~
\eta^{\rm fin}_{\Delta_\tau},\nn
\eea
where we have used that $\eta_{\Delta_\tau}\simeq 
\eta_{\Delta_\mu}$ and $\eta_{\Delta_e}\simeq 0$.
Thus in this case, it could be possible to generate a sufficient baryon 
asymmetry from the $\mu$ and $\tau$ channels even for efficiency factors
$\chi^\tau_2$ and $ ~\chi^\mu_2$ of order 0.15 and $M_2 \simeq   
10^{10}~{\rm GeV}$. For smaller efficiency factors we could still have 
successful leptogenesis with a somewhat heavier $M_2$. This situation is 
typically found for instance  in models with non-Abelian flavour symmetries. 
In this paper we have only presented a simple estimate to show that this 
mechanism can generate a large enough baryon asymmetry. A full computation 
in a model with an $SU(3)$ flavour symmetry and spontaneous $CP$ violation 
is now in progress \cite{next,su3,double}.

From \eq{washout} and taking into account that $\Delta_{11} = \sum_k W_{1k}^2
m_{\nu_k}$ we can see the washout by $N_1$ can only be stronger than the result
obtained in \eq{WeqU} if the contribution from the atmospheric neutrino mass
is sizable (assuming that $m_1 \ll m_{\rm sol}$). This requires basically
that $W_{13} m_3 \geq W_{12} m_2$. Clearly this possibility can not be
excluded, but taking into account that $m_2/m_3 \simeq 1/6$, 
$W = V_{\rm L} \cdot U$ and that 
$U_{12} \simeq 1/\sqrt{2}$ and $U_{13} \leq 0.2$ it requires large 
left-handed mixings in the neutrino Yukawa matrix.  This could be possible
in models based in Abelian flavour symmetries or discrete symmetries, although
it is clear the final result is highly model dependent and the different 
models must be studied in detail.
  
We have shown that taking correctly into account the lepton flavour dependence 
of the $CP$ asymmetries and washout processes, it is possible to obtain
successful thermal leptogenesis from the decays of the second right-handed
neutrino. The asymmetries in the muon and tau-flavour channels are then not
erased by the inverse decays of the lightest right-handed neutrino, $N_1$.
Therefore ``thermal leptogenesis'' is still viable in models 
with a strong hierarchy in the right-handed Majorana masses that is typically
the case in models with up-quark--neutrino Yukawa unification. 

\noindent
\textit{Acknowledgments:}
I acknowledge support from the RTN European project MRTN-CT-2004-503369 and 
from the Spanish  MCYT FPA2002-00612.
I thank L. Covi, S. Huber, A. Riotto and specially S. Davidson for 
very helpful discussions.

\end{document}